\documentclass{article}
\usepackage{spconf,amsmath,graphicx, comment}
\usepackage{amssymb, color}
\usepackage[noadjust]{cite}

\def\L{{\cal L}}
\def\V{{\cal V}}
\def\H{\mathcal{H}}
\def\h{\mathbf h}
\def\Y{\mathcal{Y}}
\def\B{\mathcal{B}}
\def\X{\mathcal{X}}
\def\Enc{\mathrm{Enc}}
\def\Dec{\mathrm{Dec}}
\def\CTC{\mathrm{CTC}}
\def\ASR{\mathrm{ASR}}
\def\LM{\mathrm{LM}}
\def\MLP{\mathrm{MLP}}
\def\ctc{\mathrm{ctc}}
\def\att{\mathrm{att}}

\title{Decoder-only architecture for speech recognition with \\CTC prompts and text data augmentation}
\twoauthors
  {Emiru Tsunoo, Hayato Futami, Yosuke Kashiwagi}
	{Sony Group Corporation, Japan}
  {Siddhant Arora, Shinji Watanabe}
	{Carnegie Mellon University, USA}
\begin{document}
\ninept
\maketitle
\begin{abstract}
Collecting audio–text pairs is expensive; however, it is much easier to  access text-only data.
Unless using shallow fusion, end-to-end automatic speech recognition (ASR) models require architecture modifications or additional training schemes to use text-only data.
Inspired by recent advances in decoder-only language models (LMs), such as GPT-3 and PaLM adopted for speech-processing tasks, we propose using a decoder-only architecture for ASR with simple text augmentation.
To provide audio information, encoder features compressed by CTC prediction are used as prompts for the decoder, which can be regarded as refining CTC prediction using the decoder-only model.
Because the decoder architecture is the same as an autoregressive LM, it is simple to enhance the model by leveraging external text data with LM training.
An experimental comparison using LibriSpeech and Switchboard shows that our proposed models with text augmentation training reduced word error rates from ordinary CTC by 0.3\% and 1.4\% on LibriSpeech test-clean and test-other set, respectively, and 2.9\% and 5.0\% on Switchboard and CallHome.
The proposed model had advantage on computational efficiency compared with conventional encoder--decoder ASR models with a similar parameter setup, and outperformed them on the LibriSpeech 100h and Switchboard training scenarios.
%, outperforming the conventional RNN-T and encoder--decoder ASR models of the similar parameter setup.

%Decoder-only large language models (LLMs) such as GPT3 and PaLM started to gain attraction for speech processing tasks, in which the audio information is injected as a prompt.
%However, most of the study focuses on adapting pretrained LLMs, and it is computationally costly to use in on-device applications.
%This study is the first work to train decoder-only models for speech recognition from scratch, with feasible model size for such use cases.
%We train the decoder-only transformer jointly with audio encoder and CTC, and the audio prompt is provided efficiently to the decoder by using the CTC predictions.
%Because the decoder is an architecture used for language models (LMs), these modules can be trained by targeting both automatic speech recognition (ASR) and LM tasks in parallel, so that the ASR model expands the capability of LM exploiting the external text-only data.
%Experimental comparison shows that our proposed models and training scheme performs higher accuracy than ordinary encoder--decoder ASR models with less than half the computational cost.
\end{abstract}
\begin{keywords}
Speech recognition, Decoder-only ASR, CTC, Prompt
\end{keywords}
\section{Introduction}
\label{sec:intro}

End-to-end automatic speech recognition (ASR) requires a large amount of audio–text pair data, which is difficult to obtain, whereas a large amount of text-only data can be obtained much more easily.
%Approaches to using unpaired text data for ASR tasks are not trivial.
How to exploit unpaired text data is not a trivial problem for ASR tasks.
It is common to perform score interpolation with an external language model (LM) trained with text data with an ASR model \cite{hannun2014deep,chorowski15}.
Some studies have revealed that estimating the linguistic bias in ASR models, an internal LM, is more suitable for further effective fusion \cite{meng2021ilme, zeyer2021librispeech, tsunoo22_interspeech}.
To exploit the accessible text-only data directly to the models, some studies modified the model architecture to accept both audio–text pairs and text-only data as the input for training in a multitask learning manner \cite{renduchintala2018multi, xu2020independent, wang21t_interspeech, chen22r_interspeech}.

Recently, such text resources have been effectively used by introducing pretrained large LMs (LLMs) % has shown remarkable achievements on various natural language benchmarks, 
including %encoder-only LMs such as BERT \cite{devlin2019bert} and 
decoder-only LMs such as GPT-3 \cite{brown2020language} and PaLM \cite{chowdhery2022palm} into the ASR formulation.
Studies have successfully adapted decoder-only LMs for speech-processing tasks \cite{chang2023speechprompt,zhang2023speechgpt,rubenstein2023audiopalm,fathullah2023prompting,wu2023decoder, arora23_interspeech}.
To bridge the audio--text modalities, audio information is injected into the LLMs as a prompt, which are discrete audio units  \cite{chang2023speechprompt, zhang2023speechgpt, rubenstein2023audiopalm} or continuous representations injected directly into the linguistic embedding space \cite{fathullah2023prompting, wu2023decoder}.
In the latter approach, encoded audio features are compressed by convolution layers \cite{fathullah2023prompting, wu2023decoder} or by CTC predictions \cite{wu2023decoder}, which have also been introduced in speech translation (ST) \cite{gaido2021ctc} and RNN transducer (RNN-T) decoding \cite{tian21_interspeech, wang2023accelerating}.
Owing to the strong potential of LLMs, these methods perform well in multi-task scenarios of speech-to-text processing, such as ASR, speech synthesis, and ST.
% However, compared to the ordinary ASR models, LLMs are much larger and not feasible in on-device use cases.
% Thus, we rather ask a question: "Can we build a simple ASR model with the advantage of decoder-only architecture having the strong LM ability?"
Inspired by this, we adopt a decoder-only architecture for ASR tasks that can be effectively enhanced with external text-only data.

This study aims to build a decoder-only architecture for ASR tasks and enhance its performance with text augmentation using external text-only data.
We provide audio information as prompts compressed by CTC prediction, which can also be regarded as refining CTC prediction using a decoder-only model.
Although Wu {\it et al.} attempted to train a decoder-only from scratch, they obtained a slightly degraded ST model compared to conventional encoder--decoder models.
However, we train the decoder for ASR tasks using not only the audio–text pair but also external text-only data as augmentation.
Thus, the model is trained from scratch for ASR and LM tasks simultaneously.
%Experimentally, we confirmed that the proposed model achieves better accuracy than the CTC \cite{graves06}, RNN-T \cite{graves13rnnt}, and encoder--decoder models \cite{guo2020recent} with a similar parameter setup.
Experimentally, we confirmed that the proposed model successfully refined the CTC results while achieving faster inference owing to the compression mechanism.
Using LibriSpeech 100h subset and Switchboard with text augmentation, the decoder-only model outperformed conventional encoder--decoder models with a similar parameter setup.
The main contributions of this study are as follows:
\begin{itemize}
\setlength{\leftskip}{-0.4cm}
\item We propose a new training scheme to build a decoder-only model from scratch for ASR tasks simultaneously augmented by text-only data.
\item To the best of our knowledge, this is the first study to successfully outperform conventional encoder--decoder models with the decoder-only model having similar parameter sizes by exploiting external text-only data effectively.
%\item We experimentally show that our proposed approach reduces 2.6\% and 6.0\% word error rate (WER) from the baseline CTC for LibriSpeech test-clean and test-other, respectively, using pair data of 100h with text augmentation of 960h transcription data, which outperform encoder--decoder model of the same parameter setup by 0.3\% and 1.4\% with about half the computational cost.
\item We experimentally show that our proposed approach achieves lower word error rates (WERs) by 0.3\% and 1.4\% for LibriSpeech test-clean and test-other, respectively, than the encoder--decoder model, with approximately half the computational cost, when we used paired data of 100h with text augmentation of 960h transcription data.
\end{itemize}

\begin{comment}
   This study is the first to build a simple ASR model using decoder-only architecture from scratch.
Among the aforementioned decoder-only models, only \cite{fathullah2023prompting} focused on ASR tasks.
Although they achieved superior performance in multi-lingual scenarios thanks to the LLM, the prompt was simple downsampling of the output of an encoder.
While Wu {\it et al.} attempted to train decoder-only from scratch, they ended up with slightly degraded ST model compared to conventional encoder--decoder models \cite{wu2023decoder}.
Contributions of this paper are as follows.
\begin{itemize}
\item We use an ASR model based on a decoder-only architecture with an efficient audio prompt module based on CTC prediction.
\item We propose a new training scheme to train the proposed model with both ASR and LM tasks in parallel from scratch.
\item This study is the first to successfully outperform conventional encoder--decoder models with the compact decoder-only model trained from scratch, by exploiting external text-only data effectively.
\item We experimentally show that our proposed method achieves 0.5\% and 0.9\% word error rate (WER) reduction for test-clean/other respectively using librispeech pair data of 100h and text-only data from 960h training set, with about half the computational cost.
\end{itemize} 
\end{comment}

\section{CTC prompts for decoder-only ASR}
\label{sec:ctcprompt}
\subsection{Decoder-only architecture}
\label{ssec:modelarch}
We follow the encoder--decoder conformer ASR model \cite{guo2020recent}, except that there are no source--target attention layers in the decoder transformer.
A transformer decoder that does not have source--target attention layers is considered an autoregressive decoder-only LM as GPT-3 \cite{brown2020language} or PaLM \cite{chowdhery2022palm}.

\begin{figure}
    \centering
    \includegraphics[width=1.0\columnwidth]{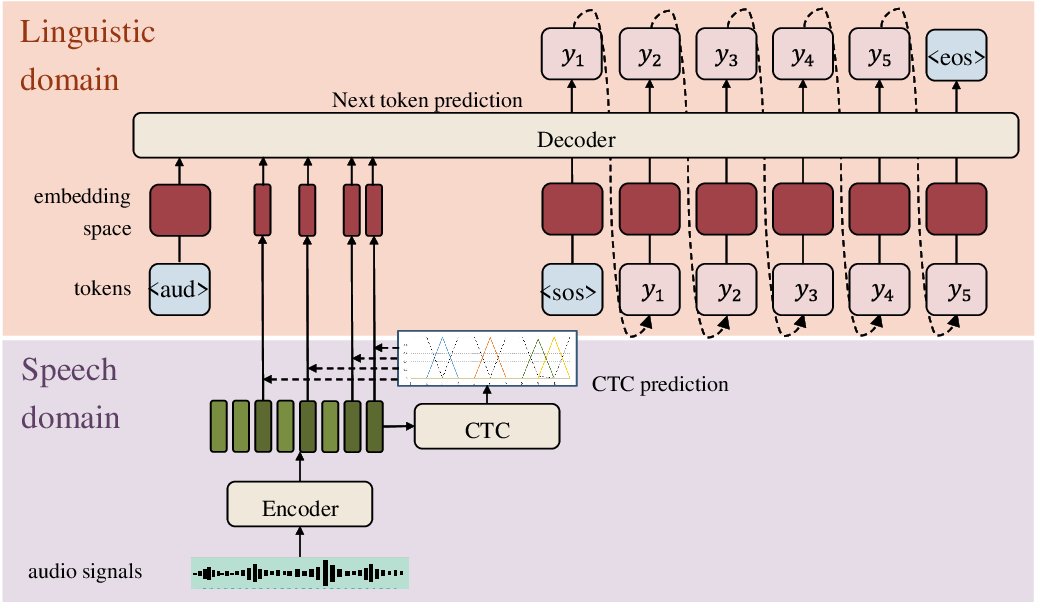}
    \vspace{-0.5cm}
    \caption{Model architecture of decoder-only model for ASR with CTC prompts.}
    \label{fig:overview}
\end{figure}

ASR is a task to predict the most probable $I$-length token sequence $\Y^I$ given a $T$-length input audio $\X^T$, i.e., $\Y^I$ with the highest probability $p(\Y^I|\X^T)$.
Instead of directly using audio input $\X^T$ in the decoder, it is generally approximated by compact audio representation. 
The audio information is provided as $\tau$-length prompts $\hat{\H}^{\tau} = \{\hat{h}_{t}|1 \leq t \leq \tau\}$ directly in the embedding space of the decoder.
We propose using CTC prediction to efficiently generate the prompts, namely, CTC prompts, which are presented in Sec.\ref{ssec:ctcprompt}.
The decoder autoregressively predicts the next linguistic token $y_i$ given an audio prompt token $\langle$aud$\rangle$, CTC prompts $\hat{\H}^{\tau}$, and the previous outputs $\Y^{<i}=\{y_{j}|0\leq j < i\}$.
Thus, 
\begin{align}
    p(\Y^i) &= \prod_{j=1}^{i} p(y_j|\hat{\H}^{\tau}, \Y^{<j}) \nonumber \\
    &=  \prod_{j=1}^{i} \Dec(\hat{\H}^{\tau}, \Y^{<j}), \label{eq:dec}
\end{align}
where $y_0$ is a start-of-sequence token, $\langle$sos$\rangle$.
An overview of the proposed method is presented in Fig.~\ref{fig:overview}.

\subsection{Audio prompt using CTC prediction}
\label{ssec:ctcprompt}
%\subsubsection{Frame reduction}
%\label{sssec:framereduction}
CTC prompts are generated using an encoder.
Generally, the length of audio is longer than that of the text sequence, which sometimes becomes problematic because the decoder self-attention computation grows quadratically with the input length.
A common approach is to use convolution layers \cite{fathullah2023prompting, wu2023decoder}, which downsample the audio input to a frame rate of 80 ms or higher to match the granularity of the linguistic tokens.
However, Wu {\it et al.} showed that the CTC-based compression \cite{gaido2021ctc} is more effective than convolution layers \cite{wu2023decoder}.
In the literature, they compared removing frames that CTC predicts as ``blank'' and averaging frames of the same CTC predictions.
We adopt the frame-removing approach, and all approaches were experimentally compared in Sec.~\ref{ssec:lib100}.

The conformer encoder outputs $T'$-length encoded audio features $\H^{T'} = \{\h_{t}|1 \leq t \leq T'\}$ from audio input $\X^{T}$ with a downsample rate of $N$ as
\begin{align}
    \H^{T'} = \Enc(\X^{T}), \label{eq:enc}
\end{align}
where $T'=T/N$.
Subsequently, the CTC module maps $\H^{T'}$ to the probability distribution of vocabulary $\V$ augmented with a blank token $\langle\mathrm{blank}\rangle$, which is jointly trained as in \cite{watanabe17}.
\begin{align}
    p(a_t) = \CTC(\h_t), \label{eq:ctc}
\end{align}
where $a_t\in \{\V \cup \langle\mathrm{blank}\rangle\}$.

%Audio prompts $\hat{\H}^{\tau} = \{\hat{h}_{t}|1 \leq t \leq \tau\}$ for the decoder is generated by the encoder.
Since the length of audio features $T'$ is generally longer than that of text sequences, audio prompts $\hat{\H}^{\tau}$ are downsampled so that $\tau < T'$.
%This can avoid computational expansion because the self-attention layers requires quadratic cost to the input length.
%The compressed audio prompt $\hat{\H}^{\tau}$ can be written as
We remove frames predicted as blank by the CTC and map them to the embedding space of the decoder, as follows.
\begin{align}
    \hat{\H}^{\tau} = \{\MLP(\h_{t})|t:\hat{a}_t\neq \langle\mathrm{blank}\rangle\},\label{eq:prompt}
\end{align}
where $\hat{a}_t$ is the most probable one in the probability distribution (\ref{eq:ctc}) and $\MLP(\cdot)$ is a mapping function from the encoder output to the embedding space, for which we adopt a linear layer for simplicity.
Thus, the compressed audio prompt is directly injected into the continuous embedding space of the decoder.
The decoder can also be seen as a refinement of the CTC prediction.

\begin{comment}
    \subsubsection{Swapped position encoding}
\label{sssec:posenc}
To distinguish audio prompt from linguistic tokens, we not only introduce audio prompt token $\langle$aud$\rangle$ preceding the audio prompt, but also use swapped position encoding for the prompt.
Conventionally the sinusoidal absolute position encoding is used in transformer architecture \cite{vaswani17} as
\begin{align}
    \mathrm{PE}_{(pos,2i)} = 
         \sin{(\frac{pos}{10000^{2i/d}})} \nonumber \\
    \mathrm{PE}_{(pos,2i+1)} = 
         \cos{(\frac{pos}{10000^{2i/d}})},
\end{align}
where $d$ is the dimension of the embedding space.
We swap $\sin$ and $\cos$ for the audio prompt to highlight the difference over the linguistic embeddings.
\end{comment}

\subsection{Training with text augmentation}
\label{ssec:training}
While ASR training requires audio--text pair data, a large text corpus is much easier to collect.
Therefore, it is reasonable to enhance ASR performance with such text data, generally with a shallow fusion of external LM models \cite{hannun2014deep, chorowski15}; for example, the LibriSpeech corpus \cite{panayotov15} provides significantly more text-only data for training external LMs in addition to the 960h audio--transcription pair data.
We focus on these scenarios and aim to enhance the decoder-only architecture using an additional text set.
%Since the architecture of the decoder is identical to a transformer LM, while the entire model is trained with an ASR task, the decoder can be trained with LM tasks seamlessly using the text data.
For the ASR task, the encoder for CTC prompts and the autoregressive decoder are jointly trained. 
Because the architecture of the proposed decoder-only model and general autoregressive LMs are represented by the same autoregressive transformer, we use text-only data for text augmentation to train only the decoder for an LM task.

The encoder and the decoder can be pre-trained using the pair data and unpaired text data, respectively, followed by fine-tuning with paired data as in \cite{fathullah2023prompting}.
However, we aimed to train them from scratch because it is considered to converge to more optimal models, which was confirmed in Sec.~\ref{ssec:lib100}.
To conduct combined training, we split the mini-batch $\B$ for both the tasks and compute each loss simultaneously, that is, $\B=\{\B_{\ASR}, \B_{\LM}\}$.

\begin{figure}
    \centering
    \includegraphics[width=0.9\columnwidth]{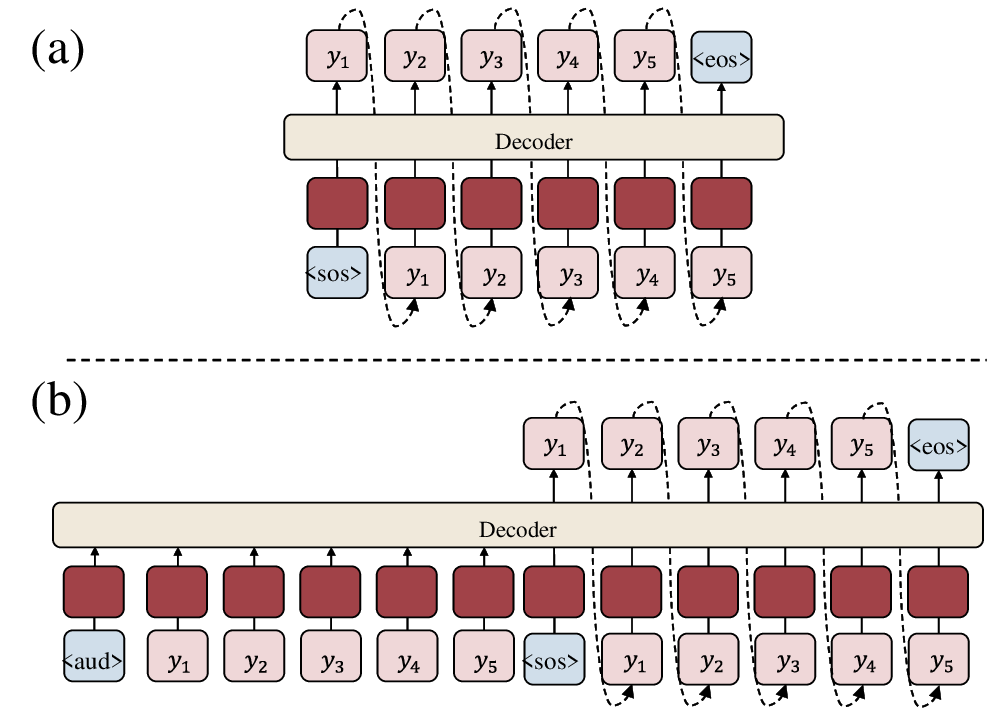}
    \vspace{-0.3cm}
    \caption{LM training methods. (a) is ordinary LM training. (b) is a pseudo prompt with the ground truth embedding vectors.}
    \label{fig:lmtraining}
\end{figure}

\begin{table*}[t]
  \caption{ASR results with LibriSpeech 100h audio--text pair data and 960h text data results. External LM was trained using external text data.}
  \label{tab:lib100}
  %\vspace{1mm}
  %\vspace{-0.3cm}
  %\centering
%  \begin{tabular}{l|cc|cc}
  \hspace{-0.4cm}
  \scalebox{0.9}{
  \begin{tabular}{c|l|c|c|cc|cccc|c|c}
    \hline
    IDs&Models&Training Data& Prompt & \multicolumn{2}{c|}{Decoding Fusion} & \multicolumn{4}{c|}{WER} & Param& RTF \\
    &&  ext-text & & CTC & ext-LM & dev-clean & dev-other & test-clean & test-other & &\\
    \hline\hline
    B1 &Baseline CTC \cite{graves06} & &&& & 9.2 & 22.8 & 9.6 & 23.5 & 34.8M & 0.14\\
    B2 &Baseline CTC \cite{graves06} & \checkmark&&& \checkmark& 7.1 & 18.8 & 7.3 & 19.6 & 44.2M & 0.29 \\
    %CTC w/ text-aug & \checkmark & \checkmark & \checkmark & & 9.0 & 21.9 & 9.3 & 22.4 & &0.14\\
    % Baseline CTC \cite{graves06} & \checkmark & & & & & 9.0 & 21.9 & 9.3 & 22.4 & 34.8M & 0.14 \\
    B3 &Baseline RNN-T \cite{graves13rnnt} & \checkmark& & &\checkmark & 7.0 & 18.6 & 7.5 & 19.4 & 54.4M & 0.67\\ %45.0M & 0.56\\
    B4 &Baseline EncDec \cite{guo2020recent} & & & & & 8.6 & 20.9 & 8.9 & 21.5 & 46.8M & 0.53 \\
    % & \checkmark &  & \checkmark & & 7.5 & 19.9 & 7.8 & 20.5 & 46.8M & 0.53 \\
    B5 & & \checkmark & & \checkmark & \checkmark & 6.9 & 18.0 & 7.3 & 18.9 & 56.2M & 0.58 \\
    \hline
    % CTC only & \checkmark &  & & & 10.3& 24.3 & 10.7 & 25.0 &  &\\
    P1 &Dec-only &  & downsample & & & 14.1 & 26.9 & 14.8 & 27.7 & 45.3M & 0.38 \\
    P2 & &  & CTC average & & & 30.5& 41.8 & 30.2& 41.2 & 45.3M &  0.55\\
    P3 & &  & CTC remove & & & 9.3& 22.4 & 9.5 & 23.0 & 45.3M & 0.23 \\
    D1 &Dec-only w/ textAug & \checkmark & CTC remove & & & 8.7& 19.6 & 9.3 & 19.9 & 45.3M&0.24 \\
    D2 && \checkmark & CTC remove & \checkmark & & 7.0 & 18.1 & 7.2 & 18.5 & 45.3M & 0.24 \\
    D3 && \checkmark & CTC remove & \checkmark & \checkmark & {\bf 6.7} & {\bf 16.8} & {\bf 7.0} & {\bf 17.5} & 54.7M & 0.29 \\
    %Dec-only w/ only (\ref{eq:lm}) & \checkmark & \checkmark & CTC remove & \checkmark& \checkmark& &  & & & 45.3M& \\
    F1 &Dec-only (fine-tuned) \cite{fathullah2023prompting} & \checkmark & CTC remove & \checkmark & \checkmark & 7.0 & 18.0 & 7.3 & 18.7 & 54.7M & 0.29 \\
    \hline
  \end{tabular}
  }
  \vspace{-0.4cm}
\end{table*}
\subsubsection{ASR training}
\label{sssec:asr}
Following the joint training of the encoder--decoder and CTC in \cite{watanabe17}, we train the encoder, decoder, and CTC simultaneously.
In each training step, in addition to the CTC loss $\L_{\ctc}$, the decoder loss $\L_{\att}$ is considered only for the transcription part, not including prompts, using teacher-forcing as:
\begin{align}
    \L_{\ASR}&=\lambda \L_{\ctc}+(1-\lambda) \L_{\att} \nonumber\\
    &= \lambda \L_{\ctc} + (1-\lambda) \sum_{i=0}^{I} - \log p(y_i|\hat{\H}^{\tau}, \tilde{\Y}^{<i}), \label{eq:asr}
\end{align}
where $\tilde{\Y}^{<i}$ comes from the ground truth and $\hat{\H}^{\tau}$ is provided by the CTC prediction of the current model introduced in (\ref{eq:prompt}).
The gradients are passed through $\hat{\H}^{\tau}$ such that the encoder parameters are updated based on (\ref{eq:asr}).
%While the CTC is not mature, because it cannot provide reasonable prompts $\hat{\H}^{\tau}$, the entire batch is used for LM training, i.e., $\B=\B_{\LM}$.
At the beginning of training, the CTC predictions are not sufficiently mature and tend to have fewer blank tokens.
This results in longer CTC prompts, which can be problematic for training purposes.
Therefore, we use tunable threshold $\theta$ to detect immature predictions.
When $\tau > \theta I$, we consider the CTC predictions to be immature and, for the particular training sentence, we replace $\L_{\att}$ with $\L_{\LM}$, which will be introduced in Sec.~\ref{sssec:lm}.

\subsubsection{LM training}
\label{sssec:lm}
Because the architecture of the decoder-only model and an LM are represented by the same autoregressive transformer, we can only train the decoder with an LM task using randomly selected text sentences from the augmented data.
A simple method is to follow ordinary LM training and minimize the negative log-likelihood, which is equivalent to minimizing the perplexity.
\begin{align}
    \L_{\LM} = \sum_{i=0}^{I} - \log p(y_i|\tilde{\Y}^{<i}) \label{eq:lm}
\end{align}
This process is illustrated in Fig.~\ref{fig:lmtraining}(a).
However, in ASR tasks, the decoder always expects CTC prompts.
Therefore, it is reasonable to minimize the gap between LM training and ASR inference by creating pseudo audio prompts with the ground-truth embedding sequence as in Fig.~\ref{fig:lmtraining}(b).
%Let the embedding sequence given by ground truth $\tilde{\Y}^{I}$ be $\H^{I}_{\tilde{\Y}}$, then the loss can be written as 
For a subset of mini-batches $\B^{*}_{\LM}$, the CTC prompts in (\ref{eq:dec}) are replaced by $\tilde{\Y}^I$, as
\begin{align}
    \L_{\LM}^{*} = \sum_{i=0}^{I} - \log 
    p(y_i|\tilde{\Y}^I, \tilde{\Y}^{<i}).
    %p(y_i|\H^{I}_{\tilde{\Y}}, \tilde{\Y}^{<i}).
\end{align}
The latter task is much easier because it copies the prompt sequence and outputs it directly.
Therefore, we split batch $\B_{\LM}$ into batches with and without pseudo-CTC prompts to guarantee the token prediction ability based on (\ref{eq:lm}), that is, $\B_{\LM} \rightarrow \{\B_{\LM}, \B^{*}_{\LM}\}$.

\section{Experiments}
\label{sec:experiment}
\subsection{Experimental setup}
\label{ssec:setup}
To evaluate the proposed decoder-only ASR model with CTC prompts, we used LibriSpeech \cite{panayotov15}, which contains an external text-only LM corpus.
We also used the Switchboard dataset, which is generally combined with the Fisher corpus, as the external text-only data for LM.
Following \cite{guo2020recent}, we trained both the baseline conformer model (Baseline EncDec) and the decoder-only model with CTC prompts (Decoder-only); the difference was that the latter did not have source--target attention layers as described in Sec.~\ref{ssec:modelarch}.
The input acoustic features were 80-dimensional filter-bank features.
The decoder (Eq.~(\ref{eq:dec})) was a 6-block transformer, and the encoder (Eq.~(\ref{eq:enc})) was a 12-block conformer, both with four-head 256-unit attention layers and 2048-unit feed-forward layers.
The models were trained using multi-task learning with CTC loss, as described in Sec~\ref{sssec:asr}, with a weight of $\lambda=0.3$ in (\ref{eq:asr}).
Because we empirically found that ASR training (Eq.~(\ref{eq:asr})) was more important than LM training with text augmentation (Eq.~(\ref{eq:lm})) for ASR tasks, we allocated most of the mini-batches to the ASR tasks and used 10\% of them for the LM tasks.
%We split the training batch into $|\B_{\ASR}|:|\B_{\LM}|:|\B^{*}_{\LM}|=18:1:1$ for training the decoder-only ASR.
Furthermore, we split the minibatch for LM tasks as $|\B_{\LM}|:|\B^{*}_{\LM}|=1:1$ as discussed in Sec.~\ref{sssec:lm}.
Threshold $\theta$ in Sec.~\ref{sssec:asr} was set to two.
We used the Adam optimizer with Noam learning rate decay.

The external LMs were trained using the text-only data, including an additional text corpus from LibriSpeech or Fisher. %for the Switchboard setup.
The LMs were two-layer LSTMs with 512 units.
We applied byte-pair encoding subword tokenization with 5,000 token classes for LibriSpeech and 2000 for Switchboard.
% In the inference, external LMs were fused with the Baseline CTC, RNN-T, EncDec, and the proposed Decoder-only models.

For comparison, we evaluated the WERs of CTC \cite{graves06} and RNN-T \cite{graves13rnnt}.
The baseline CTC used the same architecture as the aforementioned conformer encoder.
RNN-T used a 15-block conformer encoder with a one-layer 256-unit LSTM for the prediction network.
While the baseline EncDec and the proposed Decoder-only models were decoded in a label-synchronous manner, the CTC and RNN-T models used time-synchronous beam search with a beam size of 10, an LM-fusion weight of 0.4, and a length penalty of 1.0.

\begin{table*}[t]
    \centering
    \caption{An example of transcription refinement by the decoder-only architecture.}
    \label{tab:example}
    \begin{tabular}{l|l}
        \hline
         Reference &  the life of every man in the castle shall answer it if a hair of his head be signed show me his chamber \\
         CTC & the life of every man in the {\color{red} council} shall answer it if a hair of his head be {\color{red} singinged} show me his chamber \\
         Decoder-only & the life of every man in the {\bf castle} shall answer it if a hair of his head be {\bf signed} show me his chamber \\
         \hline
    \end{tabular}
      \vspace{-0.4cm}
\end{table*}

\begin{table}[t]
\caption{ASR results with LibriSpeech 960h paired data and external text results. The external LM was fused to all the models.}
\label{tab:lib960}
    \centering
    \begin{tabular}{l|cc|c}
     \hline
    Models & \multicolumn{2}{c|}{WER}&RTF\\
    & test-clean & test-other &  \\
    \hline\hline
         %Baseline CTC \cite{graves06} & 3.0 & 7.2 & 0.33\\
         Baseline CTC \cite{graves06} & 3.1 & 7.0 & 0.30\\
         Baseline EncDec \cite{guo2020recent}& {\bf 2.6} & {\bf 6.2} & 0.54\\
         \hline
         Decoder-only w/ textAug & 3.0 & 6.6 & 0.29 \\
         \hline
    \end{tabular}
      \vspace{-0.4cm}
\end{table}

\subsection{LibriSpeech 100h experiments}
\label{ssec:lib100}
First, we used the 100h clean subset of the LibriSpeech pair data and compared our proposed model under various conditions.
When using the external text-only data for the text augmentation or LM training, the transcriptions of the 960h full training set were used.
We trained the Baseline EncDec conformer, RNN-T, and the proposed decoder-only model using CTC prompts for 100 epochs.
For the Baseline EncDec and proposed Decoder-only models, we fused CTC and LM with weights of 0.4 and 0.6, respectively, and the beam size was 10.
To evaluate the pure performance of decoder-only models, we also trained the models with only the audio--text pair of 100h set, i.e., $\B=\B_{\ASR}$, without any fusion.
Furthermore, we measured the real-time factor (RTF) of the inference on the test-clean set using an 8 core 3.60 GHz Intel i9-9900K CPU.

The results are listed in Table~\ref{tab:lib100}.
First, the prompt compression methods discussed in Sec.~\ref{ssec:ctcprompt} were compared (P1--3).
Following \cite{fathullah2023prompting}, we applied convolution layers to downsample the frame rate of the encoder to 80 ms (P1).
Although they reported that a frame rate of 80 ms performed best in a multilingual evaluation, it did not perform well in our setup.
\cite{wu2023decoder} reported that averaging the frames of the same CTC predictions slightly outperformed removing blank frames for fine-tuning in the ST tasks; however, they did not use it for the models trained from scratch.
We also found that the averaging approach (P2) struggled in training for ASR tasks.
Thus, we confirmed that removing frames of blank predictions was the best choice for compressing the prompts (P3), and we used this method for the following experiments.
However, with only audio--text pair data (P3), the decoder-only architecture could not outperform the Baseline EncDec results (B4), as \cite{wu2023decoder} reported that training the decoder-only model from scratch was slightly worse than the conventional encoder--decoder model.

Next, we trained models with proposed text augmentation (D1--3).
In contrast to pair-data-only training, text augmentation using an external text-only corpus (D1) significantly reduced the WERs from 9.5\% and 23.0\% to 9.3\% and 19.9\% for the test-clean and test-other sets, respectively, compared to (P3).
As discussed in Sec.~\ref{ssec:training}, because the structure of decoder is identical to autoregressive LMs, it is an advantage of the decoder-only architecture to enhance the model with LM training without any modification.
Compared to the Baseline EncDec fused with CTC and LM (B5), the proposed method with only CTC fusion (D2) achieved comparable WERs.
The prompt compression contributed to speeding up the computation; it was less than half of the RTF of the Baseline EncDec (B5)\footnote{The number of frames was reduced down to 14.55\% on average.}.
The external LM is also effective and complementary to the text augmentation, as fusing both CTC and LM (D3) achieved the best performance; 7.0\% and 17.5\% for test-clean and test-other.
It was also comparably fast with the Baseline CTC with LM fusion (B2), in which LM computation was costly in time-synchronous decoding.
As in \cite{fathullah2023prompting}, we also evaluated the pre-training of the CTC and decoder with the paired data and unpaired text data, respectively, followed by fine-tuning using the paired data (F1), as described in Sec.~\ref{ssec:training}.
However, the fine-tuning approach was slightly worse than training from scratch (D3), presumably because the fine-tuning approach was less optimal.
Thus, using text augmentation, this study successfully outperforms the conventional encoder--decoder model with a smaller parameter size, lower RTF, and lower WERs by training the decoder-only model using same amount of data from scratch.

\subsection{LibriSpeech 960h experiments}
\label{ssec:lib960}
We also evaluated a full 960h pair set of LibriSpeech and its external text-only corpus.
We trained the models for 30 epochs and used the same beam size and fusion weights as in Sec.~\ref{ssec:lib100} for inference.
The results are listed in Table~\ref{tab:lib960}.
The decoder-only model successfully refined CTC transcription, particularly in the test-other set, from 7.0\% to 6.6\%.
We sampled an utterance from the test-other set to show how the decoder refines CTC predictions in Table~\ref{tab:example}.
However, compared with the Baseline EncDec, our method did not reach its performance.
We assume that the capacity of the decoder is relatively small because it is a 6-block transformer while a transformer LM generally has 12 blocks or more \cite{irie2019language}.
However, our proposed method still showed an advantage in the computational speed; 0.29 of RTF against 0.54.

\subsection{Switchboard and Fisher experiments}
\label{ssec:swbd}
We evaluated the Switchboard models using Hub5'00 with the Switchboard and CallHome subsets.
We fused CTC and LM with weights of 0.4 and 0.4, respectively, with a beam size of 10.
Table~\ref{tab:swbd} lists the results.
We observed a similar tendency as in Section.~\ref{ssec:lib100};
the proposed model successfully outperformed the Baseline EncDec model in both the test subsets, owing to effective text augmentation.

\begin{table}[t]
\caption{ASR results with Switchboard paired data and Fisher text corpus results. The external LM was fused to all the models.}
\label{tab:swbd}
    \centering
    \begin{tabular}{l|cc|c}
     \hline
    Models& \multicolumn{2}{c|}{WER} & RTF\\
    & Switchboard & CallHome &\\
    \hline\hline
         %Baseline CTC \cite{graves06} & 10.3& 18.7& 0.17 \\
         Baseline CTC \cite{graves06} & 8.9& 15.5 & 0.28 \\
         Baseline EncDec \cite{guo2020recent}& 8.1 & 14.8 & 0.42\\
         %Baseline EncDec \cite{guo2020recent}& 8.2 & 14.9 & 0.29 \\
         \hline
         Decoder-only w/ textAug & {\bf 7.3} & {\bf 13.3} &0.28 \\
         % Decoder-only w/ textAug & 7.4 & 13.9 &0.22 \\ beam5
         \hline
    \end{tabular}
      \vspace{-0.4cm}
\end{table}

\section{Conclusion}
\label{sec:conclusion}
We proposed a decoder-only architecture for ASR tasks that effectively applies text augmentation using text-only additional data.
Audio information was provided as CTC prompts, which mapped the output of the encoder module compressed by CTC predictions to the embedding space of the decoder.
Although the model was trained using an ASR task, the decoder was simultaneously trained for an LM task using augmented text data.
We experimentally confirmed that the proposed methods refined the baseline CTC using the decoder and outperformed conventional RNN-T and encoder--decoder models with approximately half the computational cost in LibriSpeech 100h and Switchboard setups.

Future work will include scaling up the decoder to have more capacity for LM while maintaining the advantage in inference speed. 
Extending the streaming approach can also be considered by exploiting its compactness and efficiency of the proposed approach.

\bibliographystyle{IEEEbib}
\bibliography{mybib}

\end{document}